\PassOptionsToPackage{twoside=false,a4paper,margin=25mm}{geometry}
\documentclass[nonacm]{acmart}

\AtBeginDocument{%
  }

\usepackage{colortbl}
\usepackage{pifont}
\usepackage{tabularx}
\usepackage{wasysym}
\usepackage{worldflags}
\usepackage{tabularray}
\usepackage{enumitem}
\usepackage[table]{xcolor}
\usepackage{soul}
\usepackage{tikz}
\usepackage{tcolorbox}

\newtcolorbox{recommendations}{colback=red!5!white,colframe=red!50!black,fonttitle=\bfseries,title=Recommendations}
\newtcolorbox{redbox}{colback=red!5!white,colframe=red!75!black}
\newcommand{\changed}[1]{\textcolor{black}{#1}}

\setcopyright{acmlicensed}
\copyrightyear{2026}
\acmYear{2026}
\acmDOI{XXXXXXX.XXXXXXX}
\acmConference[FAccT '26]{ACM Conference on Fairness, Accountability, and Transparency}{June ,
  2026}{Montreal, Canada}
\usepackage{fontawesome5}

\begin{document}

\newcolumntype{Y}{>{\centering\arraybackslash}X}
\newcolumntype{Z}{>{\raggedright\arraybackslash}X}

\title{\texorpdfstring{\faExclamationTriangle\ Terms of (Ab)Use: An Analysis of GenAI Services}{Terms of (Ab)Use: An Analysis of GenAI Services}}

\author{Harshvardhan J. Pandit} 
\email{me@harshp.com}
\orcid{0000-0002-5068-3714}

\affiliation{%
  \institution{AI Accountability Lab (AIAL), School of Computer Science and Statistics, Trinity College Dublin}
  \city{Dublin}
  \country{Ireland}
}

\author{Dick A. H. Blankvoort}
\email{dick.blankvoort@ru.nl}
\orcid{0009-0003-0766-4678}

\affiliation{%
  \institution{AI Accountability Lab (AIAL), School of Computer Science and Statistics, Trinity College Dublin}
  \city{Dublin}
  \country{Ireland}
}

\author{Adel Shaaban}
\email{shaabana@tcd.ie}
\orcid{0009-0009-0390-8173}

\affiliation{%
  \institution{AI Accountability Lab (AIAL), School of Computer Science and Statistics, Trinity College Dublin}
  \city{Dublin}
  \country{Ireland}
}

\author{Sasha Luccioni}
\email{sasha.luccioni@huggingface.co}
\affiliation{%
  \institution{HuggingFace}
  \city{Montreal}
  \country{Canada}
}
\orcid{0000-0001-6238-7050}

\author{Abeba Birhane}
\email{birhanea@tcd.ie}
\affiliation{%
  \institution{AI Accountability Lab (AIAL), School of Computer Science and Statistics, Trinity College Dublin}
  \city{Dublin}
  \country{Ireland}
}
\orcid{0000-0001-6319-7937}
\renewcommand{\shortauthors}{Pandit et al.}

\begin{abstract}
Generative AI services like ChatGPT and Gemini are some of the fastest-growing consumer services. Individuals using such services must accept their terms of use before access, and conform to these terms for continued use of the service. Established literature has shown that despite their status as legally-binding agreements, terms of use are not actually well-understood, and may contain implications that are surprising for consumers. In this paper, we analyse the terms of 6 generative AI services from the perspective of an EU-based consumer. 
Our findings, based on a developed codebook which we provide in the paper, reiterate known issues regarding generative AI services such as the default use of user data for training and surface new concerns regarding responsibility, liability, and rights. All terms in our analysis contained language that explicitly discards assurances regarding the quality, availability and appropriateness of the service, regardless of whether the service is free or paid. The terms also make users solely responsible for outputs meeting norms dictated by the provider, despite no information or control being provided over the functioning of the model, and at the risk of account termination. The terms further restrict users in how outputs can be used while service providers utilise both user-provided inputs as well as user-liable outputs for a wide variety of purposes at their discretion.
The implications of these practices are severe, as we find consumers suffer from lack of necessary information, significant imbalance of power, and have responsibilities they cannot materially fulfil without violating the terms. To remedy this situation, we make concrete recommendations for authorities and policymakers to urgently upgrade existing consumer protection mechanisms to tackle this growing issue.

\end{abstract}

\maketitle
\subsubsection*{\textbf{Status:} Peer-reviewed, to be presented at ACM Conference on Fairness, Accountability, and Transparency (FAccT) 2026 \\ \textbf{Project website:} see \underline{\url{https://aial.ie/research/terms-of-abuse}} for supplementary materials and updates \\ Copyright held by the owner/author(s). Publication rights licensed to ACM. Permission to share under CC-BY.}

\section{Introduction}\label{sec:intro}
\begin{quote}
\begin{redbox}
    \textit{Imagine you purchase bread for a sandwich but then discover that your toaster refuses to work because the bread is `unauthorised'. 
    You read the toaster's terms and conditions and find that it had lots of `rules' on what you could or couldn't do with the toaster, and find you were unaware of these restrictions until now, and you wonder if these are legal. In any case, you do not have your sandwich.}
\end{redbox}
\end{quote}
While this is a scenario adapted from Doctorow's `\textit{Unauthorised Bread}' \cite{doctorow2018unauthorized}, it is also a reflection on how consumer's lives are increasingly shaped by digital/online services where they accept `terms and conditions' involving rules they do not understand well \cite{obarBiggestLieInternet2020}.
These `terms and conditions' are contracts that govern the provision of digital services, and are also commonly referred to as `terms of use' or `terms of service' (henceforth just \textit{terms}).

Despite their importance, consumers have been shown to not read the terms or the privacy policies because they are lengthy, cumbersome to comprehend, and confusing \cite{robinson2020beyond,karanicolas2021too,obarBiggestLieInternet2020,litman-navarroOpinionWeRead2019}. 
In fact, `Terms and conditions' are now a meme, being defined by the Urban Dictionary as ``\textit{That long a** wall of text no one bothered reading, yet everyone agrees}''~\cite{urbandict}. 
To prevent companies from taking advantage of consumers through the terms, consumer protection laws restrict the content and use of terms by requiring transparency, such as the provider's identity \cite{EU_consumer-rights-directive}, while prohibiting terms and practices deemed unfair \cite{EU_UCTD,EU_UCPD,degeestSigningWithoutReadingProblemAnalysis2002,loosWantedBiggerStick2016}.
Yet, consumer protection laws continue to rely on the terms to dictate expectations regarding quality, suitability, and responsibilities (liabilities), making them indispensable in shaping consumer's experience in the use of digital services.

The rapid advances in generative AI (GenAI) technologies have led to the proliferation of a new wave of digital services powered by large language models (LLMs)\changed{, and their direct regulation through the AI Act \cite{EU_AIAct}}.
Amongst these, big tech corporations such as Alphabet, Amazon, Apple, Meta, and Microsoft as well as relatively new entrants such as OpenAI and DeepSeek have emerged as dominant developers and service providers. 
The extent of GenAI uptake has been so rapid that OpenAI's ChatGPT currently ranks amongst the top 5 websites visited as well as amongst the top apps in both Apple and Google app stores, and is projected to have over 700 million users.\footnote{Web rankings from Similarweb and Semrush, app store rankings from iOS App Store and Google Play Store, users reported from OpenAI's website.}
Several critical and urgent questions are being raised about GenAI, in particular regarding their legality, ethics, and fairness due to requiring huge volumes of data to train and fine-tune large language models (LLMs) \cite{birhane2023into,luccioni2021s} as well as due to concerns rooted in privacy and user agency \cite{bender2021dangers,brown2022does}.
What has received comparatively lesser attention is how the terms of GenAI services shape the experiences of consumers, in particular whether they repeat the known challenges highlighted earlier and whether they create new forms of `unfairness' that warrant further investigations and changes to consumer protection legislation. \changed{Though we focus here on the EU, this topic and our work has a broader appeal for other jurisdictions based on the similarities of the terms and the resulting consumer experiences.}

In this paper, we read and manually annotated the terms of six prominent web-based GenAI services with a focus on EU as the jurisdiction. 
Our findings reveal significant issues regarding structure, accessibility, lack of transparency, asymmetry in freedoms and responsibilities, and jurisdictional rights.
Following insights from our analysis, we argue that the current practices and terms as used by GenAI services are likely to not be in good faith and are imbalanced, thereby making them potentially unfair for consumers.
We also present avenues for specific further analysis and investigations in order to answer raised questions, and identify concrete recommendations to inform consumer protection authorities and current regulatory proposals.
\changed{
In particular, our work extends prior efforts related to analysis of privacy policies [20,22], arguments about unfairness [17], issues reading terms [19,23,26], and consumer contracts [24]. To the best of our knowledge, this work is thus the first to analyse GenAI terms for unfairness -- including AI-specific analysis --and relate this to EU consumer protection laws, while providing our codebook and findings for reuse/extensibility. While many issues we highlight are also applicable to general service terms, our focus is on generative AI services as it is rapidly shaping and outpacing policy, and thus warrants dedicated investigation.
Altogether, we make three major contributions:}
\begin{enumerate}
    \item We provide a methodological framework for performing a fine-grained \changed{annotation and assessment of the consumer terms used by GenAI services;}
    \item We manually annotate the terms of use of six GenAI services and present timely insights that highlight failures regarding `good faith', lack of transparency, and imbalances as defined in EU consumer protection laws;
    \item We provide concrete recommendations for policy makers to both interpret existing consumer protection regulation as well as shape new laws, to ensure better consumer protection and fairness in GenAI services.
\end{enumerate}

The rest of the paper is organised as follows:
Section~\ref{sec:sota} provides an overview of relevant work and state of the art regarding analysis of terms, including a specific focus on GenAI services' practices.
Section~\ref{sec:methods} describes our methodology and provides detailed process on how we 
developed codebook, identified the six services, and our annotation process.
Section~\ref{sec:findings} presents our findings grouped under four categories: Structure and Access, Service Provision \& Quality Assurance, Inputs \& Outputs, and Legal.
Section~\ref{sec:discuss} addresses the 
implications of these findings from the perspective of consumers and 
in context of existing consumer protection law, and Section~\ref{sec:conclusion} provides conclusions.
\changed{Further updates and resources are provided at \url{https://aial.ie/research/terms-of-abuse}.}

\section{Related Work}\label{sec:sota}

\textbf{Difficulties in understanding terms and policies by consumers:}
That terms provided by services exhibit several difficulties that prevent consumers from understanding their implications has been well established in literature.
For example, Luger et al. \cite{lugerConsentAllRevealing2013} found difficulties in terms stemming from complexity of language, McDonald \& Cranor \cite{mcdonaldCostReadingPrivacy2008} point to the time and effort required to read lengthy policies, and Obar \& Oeldorf-Hirsch \cite{obarBiggestLieInternet2020} point to the information overload faced by readers -- calling the uninformed acceptance as ``The biggest lie on the Internet''.

\noindent\textbf{Unfairness analysis under consumer protection law:}
Geest \cite{degeestSigningWithoutReadingProblemAnalysis2002} discuss the `Signing-Without-Reading Problem' and the difficulties faced by consumers regarding terms in the context of EU consumer protection laws, in particular the Unfair Contract Terms Directive \cite{EU_UCTD}. They also point to the limitations of law in dealing with emerging practices that stretch and go beyond what should be considered as `unfair', and point to terms now containing clauses that would `never be accepted by rational parties'.
Loos \& Luzak \cite{loosWantedBiggerStick2016} identified terms used by  online service providers as unlikely to pass the Directive’s unfairness test. These included Google, Facebook, and Twitter (now X) who have since become GenAI service providers. The paper points to systemic and common patterns across all analysed terms regarding unilateral changes to service by providers, unilateral termination of contract, liability limitations, ambiguity in applicable law, lack of transparency, and the imposition of specific jurisdictions. 

\noindent\textbf{Analysis of GenAI terms and policies:}
Tao et al. \cite{taoLongitudinalMeasurementPrivacy2025} conducted a longitudinal analysis of privacy policies from 11 GenAI providers from US, EU, and China, and found increases in length and vagueness over time.
King \& Klyman et al. \cite{king2025user} analysed privacy policies of six US GenAI providers regarding their data collection and training practices based primarily on the California Consumer Privacy Act (CCPA). Their findings show that providers may be training models based on data disclosed in chats, including sensitive information such as biometric and health data, and that the privacy policies themselves were insufficiently transparent regarding this. They conclude with concerns regarding the accountability of GenAI providers and recommendations for GenAI providers and policymakers.
Edwards et al. \cite{edwardsPrivateOrderingGenerative2025} analysed 13 providers terms and policies to identify implications regarding copyright and data protection rights. They found risk arising from copyright infringement, privacy breaches, and illegal content was assigned unconditionally to the user, and posit that the GenAI providers are moving towards `platformisation paradigm' where the providers are `neutral intermediaries' in the style of DMCA content moderation. They explicitly point to the need for further investigations ``if competition in the market itself produces fairer and more balanced T\&Cs or if, as seems sadly more likely, a de facto cartel continues to impose the clauses most favourable to a small handful of extremely powerful tech companies''. 
Our paper's goals and objectives thus naturally follow from this and other past work.

\section{Methodology}\label{sec:methods}

The goal of our study was to analyse the terms that GenAI services require consumers to accept, and to assess whether they are transparent and fair to users.
To implement this, we developed a codebook and used it to identify information from the terms.
We then analysed and discussed the findings to identify implications regarding consumer experiences and the law.
In this section, we explain the process and the methodologies we used to implement the study.

\subsection{Scope and selection process}
GenAI services typically allow individuals (as consumers) to use services for free (i.e. without payment) or through a paid account, and separately also allow individuals to access services as part of an enterprise account (i.e. as corporate users). For this study, we limited the scope to only analyse consumer uses of GenAI services, and excluded enterprise services and features. 
We chose two providers each for the following criteria: established big-tech actors -- Google (Gemini), Microsoft (Copilot), emergent non-US providers -- Mistral (LeChat, France), Deepseek (China), and new prominent providers -- OpenAI (ChatGPT), Anthropic (Claude).
Other services that were identified but not selected were reserved for future analysis, and include Meta, xAI, Perplexity, and Qwen, amongst others.
\changed{Of relevance, Microsoft and Google are platforms where the same account and terms also cover the other non-GenAI services. For our work, we only considered the GenAI aspects to be in scope.}

To identify relevant terms, we used the links provided in the web interface of each service.
When the linked terms document involved hyperlinks to other documents, we only included the linked documents if they were clearly mentioned as being `part of the terms'.
We specifically did not consider linked documents that were mentioned as informational or providing guidance, as our goal was to study the terms as a contractually binding provision.
Following this rule, we only included `privacy policies', `acceptable use', and similar documents only when they were referenced as being part of the terms.
\changed{For example, OpenAI's terms note their privacy policy ``does not form part of these Terms''. We also did not include branding guidelines, contact or `about us' pages, or links in headers and footers.}
\begin{table}[]
    \centering
    \caption{Summary view of codebook used to analyse GenAI service terms: Fields are grouped into categories and shown with their identifier and specific topics of enquiry. Fields for inputs and outputs are shown together using common topics where possible.}
    \label{table:codebook-overview}
    \begin{tabular}{|l|l|l|l|}
    \hline
        \textbf{ID} & \textbf{Field} & \textbf{ID} & \textbf{Field} \\ \hline
        \cellcolor{lightgray}\textbf{S1} & \cellcolor{lightgray}\textbf{Metadata} & S3.7 \& S4.7 & Model training by provider \\ 
        S1.1-S1.6 & metadata for annotator, ID, timestamp, target & S3.8 \& S4.8 & User control for training \\ 
        S1.7 & Terms structure & S3.9 \& S4.9 & Used for other purposes \\ 
        S1.8 & Terms for free/paid service & S3.10 \& S4.10 & Third Party sharing \\ 
        \cellcolor{lightgray}\textbf{S2} & \cellcolor{lightgray}\textbf{Service Provision} & S3.11 \& S4.11 & Third Party identity \\ 
        S2.1 & Service description & S3.12 \& S4.12 & Third Party sharing purposes \\ 
        S2.2 & Features and functionalities & S3.13 \& S4.13 & Third Party sharing privacy \\ 
        S2.3 & Stability of Service & S3.14 \& S4.14 & User responsibilities \\ 
        S2.4 & Quality of Service & S3.15 \& S4.15 & Liability \\ 
        S2.5 & Quality Assurances & S3.16 \& S4.16 & Use of filtering/detection \\ 
        \cellcolor{lightgray}\textbf{S3 \& S4} & \cellcolor{lightgray}\textbf{Inputs (S3) \& Outputs (S4)} & S3.17 \& S4.17 & Invalid input/output consequence \\ 
        S6.A & Common label for I/O & S6.E & Human involvement \\ 
        S6.B & I/O as personal data & S4.18 & Risks \\ 
        S3.1 \& S4.1 & Modalities & S4.19 & Model training by Consumer \\ 
        S3.2 \& S4.2 & Sources & S4.20 & Harms regarding use of service \\ 
        S3.3 \& S4.3 & Exclusively permitted I/O & \cellcolor{lightgray}\textbf{S5} & \cellcolor{lightgray}\textbf{Legal} \\ 
        S3.4 \& S4.4 & Prohibited I/O & S5.1 & Restriction of Jurisdiction \\ 
        S6.C & Reverse engineering & S5.2 & Specifically applicable laws \\ 
        S3.5 \& S4.5 & Intended service scope & S5.3 & Local jurisdiction \\ 
        S3.6 & Rights over inputs & S5.4 & Restricted arbitration \\ 
        S4.6 & Rights over outputs & S6.D & Restrictions and applicable law \\ \hline
    \end{tabular}
\end{table}
\subsection{Codebook Development}
To develop the codebook, we used a design methods approach \cite{Gericke_Eckert_Stacey_2022} where we
first read terms documents from multiple providers to familiarise ourselves with the structure, content, and format of these documents.
Based on these, we identified three high-level organising themes: service provisions, model inputs and outputs, and legal considerations. We then developed a preliminary codebook with questions under each theme for structured extraction of relevant information. We performed pilot annotations on multiple terms to assess the extent to which the preliminary codebook could capture important and relevant information to meaningfully achieve our goal as well as to identify limitations and weaknesses in our approach. 
Following pilot annotations, we deliberated and iteratively refined the codebook where some questions deemed less relevant were dropped, others further sharpened, our organising themes expanded (from three to five categories) and limitations and weaknesses were addressed to create the final codebook.

The final codebook, summarised in Table.\ref{table:codebook-overview}, contained a total of 59 questions grouped under 5 sections: Metadata, Service Provision, Inputs, Outputs, and Legal.
Unique identifiers were assigned to each category (e.g. \textit{S1} for \textit{Section 1}) and questions (e.g. \textit{S1.1} for \textit{Section 1 Question 1}) to aid in discussions and analysis.
To assist the annotators in identifying and documenting information, each question was accompanied with an enumeration of common values based on expected answers with the possibility to note additional information not covered by these. 
Each question was also accompanied with examples of relevant statements in terms to help annotators identify information more easily.
The complete codebook with the questions and enumerated values is provided in Appendix \ref{appendix:codebook}.

\subsection{Annotation Process}
To analyse the 6 identified services and their terms, we utilised 6 unique pairs of 4 annotators which resulted in 12 total annotations.
Annotators were encouraged to note the specific sentences or sections relevant to the question as part of the provided notes field in the codebook, as well as to note the correct answers where the codebook's options were not sufficient.
Annotators were specifically instructed to avoid subjective interpretations of both questions and statements and only identify information based on factual statements in the terms.
\changed{The pilot annotations were used as shared knowledge and in discussions for understanding how terms are structured and how to interpret them.}
After completion of annotations, the information identified across the two annotations for each service was compared.
If the two annotators had found the same information, this was considered as establishing consensus. 
\changed{For all other annotations, the discrepency in information was resolved in two ways. First, where only one annotator reported information while the other annotator did not, either due to missing the information or due to misinterpreting it as being relevant. This happened in 57 cases (Claude (10), Deepseek (15), Gemini (11), Copilot (15), Le Chat (2), ChatGPT (5)). Second, where both annotators reported different information. This happened in 29 cases (Claude (4), Deepseek (8), Gemini (4), Copilot (9), Le Chat (0), ChatGPT (3).}
One annotator with prior experience of working with legal terms acted as the adjudicator and consulted each pair to identify the correct information and reach consensus, including going through the terms to confirm each annotator's understanding of the information.
All discrepancies were successfully resolved through this process.

\section{Findings \& Analysis}\label{sec:findings}
In this section we describe our findings based on the annotations completed using the codebook.
We indicate the relevant questions from the codebook using their identifiers. 
For each annotation, we only used archived copies of terms and documents as found on 6th November 2025.
In total, we identified and annotated 21 documents (see 
Appendix~\ref{appendix:terms}) during November 2025, and confirmed the latest terms were used. The complete findings are provided in Appendix~\ref{appendix:findings}, and summarised findings are provided in Table~\ref{table:findings-quality} regarding service quality; Table~\ref{table:findings-users} regarding user responsibilities \& liabilities, Table~\ref{table:findings-provider} regarding provider responsibilities, and Table~\ref{table:findings-legal} regarding applicable laws.

\subsection{Service Quality}

\begin{table}[ht] 
    \centering
    \caption{Overview of information provided in GenAI service terms related to service provision, quality, and assurance. A check mark (\ding{51}) indicates the terms provided this information, and cross mark (\ding{55}) indicates they did not. Colours are used to draw attention to the severity of implications.} 
    \label{table:findings-quality}
    \small
    \begin{tabularx}{\textwidth}{|l|Y|Y|Y|Y|Y|Y|Y|}
    \hline
    \textbf{Topic} & \textbf{Claude (Anthropic)} & \textbf{DeepSeek (Deepseek)} & \textbf{Gemini (Google)} & \textbf{Copilot (Microsoft)} & \textbf{Le Chat (Mistral)} & \textbf{ChatGPT (OpenAI)} \\ \hline
    Service functionalities & \cellcolor{red!25}\ding{55} & \cellcolor{red!25}\ding{55} & \cellcolor{red!25}\ding{55} & \cellcolor{red!25}\ding{55} & \cellcolor{red!25}\ding{55} & \cellcolor{red!25}\ding{55} \\ \hline
    Performance Assurance & \cellcolor{orange!25}\ding{55} & \cellcolor{orange!25}\ding{55} & \cellcolor{orange!25}\ding{55} & \cellcolor{orange!25}\ding{55} & \cellcolor{orange!25}\ding{55} & \cellcolor{orange!25}\ding{55} \\ \hline
    Stability Assurance & \cellcolor{orange!25}\ding{55} & \cellcolor{orange!25}\ding{55} & \cellcolor{orange!25}\ding{55} & \cellcolor{orange!25}\ding{55} & \cellcolor{orange!25}\ding{55} & \cellcolor{orange!25}\ding{55} \\ \hline
    Service change without notice & \cellcolor{red!25}\ding{51} & \cellcolor{red!25}\ding{51} & \cellcolor{red!25}\ding{51} & \cellcolor{red!25}\ding{51} & \cellcolor{red!25}\ding{51} & \cellcolor{red!25}\ding{51} \\ \hline
    Accuracy Assurance & \cellcolor{orange!25}\ding{55} & \cellcolor{orange!25}\ding{55} & \cellcolor{orange!25}\ding{55} & \cellcolor{orange!25}\ding{55} & \cellcolor{orange!25}\ding{55} & \cellcolor{orange!25}\ding{55}\\ \hline
    Warranty disclaimers & \cellcolor{red!25}\ding{51} & \cellcolor{red!25}\ding{51} & \cellcolor{red!25}\ding{51} & \cellcolor{red!25}\ding{51} & \cellcolor{red!25}\ding{51} & \cellcolor{red!25}\ding{51} \\ \hline
    \end{tabularx}
\end{table}

No providers provided a clear definition or description of their service regarding specific features or functionalities being provided to the user (\textit{S2.2}) despite this being a necessary part of the terms to indicate what the service and payments entail \cite{CommissionNoticeGuidance2019}. 
All terms mentioned that the terms may change at any time and that the consumer would be given a notice period of 30 days to accept or leave (\textit{S2.3}), while also mentioning the service may change without prior notice. 
No terms described the quality of the service (\textit{S2.4}) or provided quality assurances (\textit{S2.5}), instead warranty disclaimers had explicit statements on not providing assurances regarding performance, fitness for purpose, errors, or defects.
\changed{From these, we point to lack of service functionalities, changes to service without notice, and use of warranty disclaimers are important regarding implications (indicated using red in table).}

\subsection{User Rights, Responsibilities, \& Liabilities}

\begin{table}[!ht]
    \centering
    \caption{Overview of information provided in GenAI service terms regarding user responsibilities and consequences. A check mark (\ding{51}) indicates the terms provided this information, and cross mark (\ding{55}) indicates they did not. Colours are used to draw attention to severity of implications.}
    \label{table:findings-users}
    \small
    \begin{tabularx}{\textwidth}{|p{3.5cm}|Y|Y|Y|Y|Y|Y|Y|}
    \hline
    \textbf{Topic} & \textbf{Claude (Anthropic)} & \textbf{DeepSeek (Deepseek)} & \textbf{Gemini (Google)} & \textbf{Copilot (Microsoft)} & \textbf{Le Chat (Mistral)} & \textbf{ChatGPT (OpenAI)} \\ \hline
    Input/output restrictions & \cellcolor{orange!25}\ding{51} & \cellcolor{orange!25}\ding{51} & \cellcolor{orange!25}\ding{51} & \cellcolor{orange!25}\ding{51} & \cellcolor{orange!25}\ding{51} & \cellcolor{orange!25}\ding{51} \\ \hline
    Users input liability & \cellcolor{red!25}\ding{51} & \cellcolor{red!25}\ding{51} & \cellcolor{red!25}\ding{51} & \cellcolor{red!25}\ding{51} & \cellcolor{red!25}\ding{51} & \cellcolor{red!25}\ding{51} \\ \hline
    Provider input liability & \cellcolor{orange!25}\ding{55} & \cellcolor{orange!25}\ding{55} & \cellcolor{orange!25}\ding{55} & \cellcolor{orange!25}\ding{55} & \cellcolor{orange!25}\ding{55} & \cellcolor{orange!25}\ding{55} \\ \hline
    User output liability & \cellcolor{orange!25}\ding{51} & \cellcolor{orange!25}\ding{51} & \cellcolor{orange!25}\ding{51} & \cellcolor{orange!25}\ding{51} & \cellcolor{orange!25}\ding{51} & \cellcolor{orange!25}\ding{51} \\ \hline
    Provider output liability & \cellcolor{red!25}\ding{55} & \cellcolor{red!25}\ding{55} & \cellcolor{red!25}\ding{55} & \cellcolor{red!25}\ding{55} & \cellcolor{red!25}\ding{55} & \cellcolor{red!25}\ding{55} \\ \hline
    Underlying model user access & \cellcolor{orange!25}\ding{51} & \cellcolor{orange!25}\ding{51} & \cellcolor{orange!25}\ding{51} & \cellcolor{orange!25}\ding{51} & \cellcolor{orange!25}\ding{51} & \cellcolor{orange!25}\ding{51} \\ \hline
    Guidance for responsibilities & \cellcolor{red!25}\ding{55} & \cellcolor{red!25}\ding{55} & \cellcolor{red!25}\ding{55} & \cellcolor{red!25}\ding{55} & \cellcolor{red!25}\ding{55} & \cellcolor{red!25}\ding{55} \\ \hline
    Violation causes suspension & \cellcolor{orange!25}\ding{51} & \cellcolor{orange!25}\ding{51} & \cellcolor{orange!25}\ding{51} & \cellcolor{orange!25}\ding{51} & \cellcolor{orange!25}\ding{51} & \cellcolor{orange!25}\ding{51} \\ \hline
    Violation causes termination & \cellcolor{orange!25}\ding{51} & \cellcolor{orange!25}\ding{51} & \cellcolor{orange!25}\ding{51} & \cellcolor{orange!25}\ding{51} & \cellcolor{orange!25}\ding{51} & \cellcolor{orange!25}\ding{51} \\ \hline
    \end{tabularx}
\end{table}

All terms had specific prohibitions regarding input/output (\textit{S3.4,S4.4}) which included legal prohibitions (e.g. illegal content, copyright, intellectual property rights) and safety prohibitions (e.g. producing harmful content or misinformation).
All terms except Google's also had explicit prohibitions regarding `jailbreaking' and reverse engineering (\textit{S6.C}), which would mean that users had no access to the underlying models and must utilise only the provided interfaces.
None of the terms clarified what inputs or outputs are exclusively permitted (\textit{S3.3,S4.3}) or restricted them to a specific scope (\textit{S3.5,S4.5}). 
This meant all terms only had prohibitions on what inputs and outputs were invalid, and did not indicate what the service is suitable or designed for in terms of a specific set of inputs or outputs or scenarios.

For inputs provided by users, all terms mentioned that the users had sole responsibility to ensure the inputs meet the stated requirements and prohibitions (\textit{S3.14}).
Similarly, all terms also mentioned the user was solely responsible for ensuring all outputs produced through the service met requirements and prohibitions (\textit{S4.14}). 
In addition to stating responsibility, each terms document also had a disclaimer that made the users liable for both inputs as well as outputs, in particular regarding third party grievances.
No terms provided a rationale or explanation or a link to resources for users to understand and control how their inputs resulted in a responsibility and liability over outputs, especially since the users could only access the underlying model through the provided service.
All terms mentioned consequences of violating the terms, including the defined restrictions and prohibitions, as resulting in potential suspension of the service or termination of the contract. Claude's terms had a third possibility of degrading the service, though it did not clarify what this implied.
No terms provided a clear criteria or assessment of violations regarding severity, or which specific violations would result in a suspension and which would cause termination, though all terms clarified consequences on subscription fees in case of contract termination.
\changed{From these, the users having liability for inputs while providers having no liability for outputs, and the lack of guidance to users are important finding in discussing implications (indicated using red in table).}

\subsection{Provider Benefits \& Responsibilities}

\begin{table}[!ht] 
    \centering
    \caption{Overview of information provided in GenAI service terms regarding providers use of inputs and outputs for training and other purposes, user controls, and human involvement in decision making. A check mark (\ding{51}) indicates the terms provided this information, and cross mark (\ding{55}) indicates they did not. Colours are used to draw attention to severity of implications.}
    \label{table:findings-provider}
    \small
    \begin{tabularx}{\textwidth}{|p{3.5cm}|Y|Y|Y|Y|Y|Y|Y|}
    \hline
        \textbf{Topic} & \textbf{Claude (Anthropic)} & \textbf{DeepSeek (Deepseek)} & \textbf{Gemini (Google)} & \textbf{Copilot (Microsoft)} & \textbf{Le Chat (Mistral)} & \textbf{ChatGPT (OpenAI)} \\ \hline
        Training on input/outputs & \cellcolor{red!25}\ding{51} & \cellcolor{red!25}\ding{51} & \cellcolor{gray!25}\,\,\textbf{N/A}* & \cellcolor{red!25}\ding{51}& \cellcolor{red!25}\ding{51} & \cellcolor{red!25}\ding{51} \\ \hline
        Users must opt-out & \cellcolor{red!25}\ding{51} & \cellcolor{red!25}\ding{51} & \cellcolor{red!25}\,\,\ding{51}* & \cellcolor{red!25}\ding{51} & \cellcolor{red!25}\ding{51} & \cellcolor{red!25}\ding{51} \\ \hline
        Data still used after opt-out & \cellcolor{red!25}\ding{51} & \cellcolor{green!25}\ding{55} & \cellcolor{green!25}\ding{55} & \cellcolor{green!25}\ding{55} & \cellcolor{green!25}\ding{55} & \cellcolor{green!25}\ding{55} \\ \hline
        Inputs/outputs as personal data &  \cellcolor{orange!25}\ding{51} & \cellcolor{orange!25}\ding{51} & \cellcolor{orange!25}\ding{51} & \cellcolor{orange!25}\ding{51}& \cellcolor{orange!25}\ding{51} & \cellcolor{orange!25}\ding{51} \\ \hline
        Inputs/outputs for advertising & \cellcolor{red!25}\,\,\ding{51}* & \cellcolor{red!25}\,\,\ding{51}* & \cellcolor{red!25}\,\,\ding{51}* & \cellcolor{red!25}\ding{51}& \cellcolor{red!25}\,\,\ding{51}* & \cellcolor{red!25}\,\,\ding{51}* \\ \hline
        Third-party sharing & \cellcolor{red!25}\ding{51} & \cellcolor{red!25}\ding{51} & \cellcolor{red!25}\ding{51} & \cellcolor{red!25}\ding{51} & \cellcolor{red!25}\ding{51} & \cellcolor{red!25}\ding{51} \\ \hline
        Automated filtering/detection & \cellcolor{orange!25}\ding{51} & \cellcolor{orange!25}\ding{51} & \cellcolor{orange!25}\ding{51} & \cellcolor{orange!25}\ding{51} & \cellcolor{orange!25}\ding{51} & \cellcolor{orange!25}\ding{51} \\ \hline
        Moderation involves humans & \cellcolor{red!25}\ding{55} & \cellcolor{gray!25}\textbf{N/A} & \cellcolor{red!25}\,\,\ding{55}* & \cellcolor{red!25}\ding{55} & \cellcolor{gray!25}\textbf{N/A} & \cellcolor{red!25}\ding{55} \\ \hline
    \end{tabularx}
\end{table}

All service providers except DeepSeek make no distinction between input and output data, and instead compound them under a single label (\textit{S6.A}): \textit{`Materials'} (Claude), \textit{`Content'} (Gemini, Copilot, ChatGPT), \textit{`Data'} (Le Chat). 
All service providers with the exception of DeepSeek also explicitly treat both inputs and outputs as personal data (\textit{S6.B}).
Where terms did not explicitly refer to inputs or outputs and used the common label, we interpreted this to mean that clause applies to both inputs and outputs, and similarly, where the terms mentioned personal data we interpreted this as also applying to inputs and outputs.
All providers, except Google, explicitly asserted that the user inputs and outputs would be used for training GenAI models by the provider (\textit{S3.7,S4.7}), and the only option provided to users was to opt-out (S3.8,S4.8).
This confirms earlier findings from King \& Klyman et al. \cite{king2025user} who analysed GenAI privacy policies, though they mention Google's privacy policy containing the training and opt-out declarations, which we did not analyse as it was not linked within Google's terms.
Anthropic's terms specifically stated that even if the user opts out, if the user gives feedback (e.g. by clicking the thumbs up/down buttons in chat), their conversations would be used for training and the user had no control over this.
\changed{These findings also have relevance under the GDPR as it constitutes a change from consent (opt-in) to legitimate interest (opt-out) in the use of personal data, as reiterated by \cite{king2025user}, and is of relevance to the European Data Protection Board (EDPB) opinion requiring a proportionality assessment for justifying this under the GDPR
\cite{EDPB_28/2024}.}

Copilot (Microsoft) was the only service whose terms indicated that inputs and outputs may be used for advertising, including forms that involve third-parties, tracking, and profiling.
All terms mentioned inputs and outputs being shared with third parties -- where by third party we mean any entity other than the user and the service provider (\textit{S3.10,S4.10}).
Only Mistral provided a detailed list of third parties with their roles.
All terms mentioned that inputs and outputs would also be used for other purposes beyond service provision and training, such as research and development, fraud management, and analytics (\textit{S3.9,S4.9}). 
All providers asserted that rights over provided inputs were retained by the user (\textit{S3.6}) and granted rights over produced outputs (\textit{S4.6}). 
However, all terms explicitly prohibited use of output data for training by the consumer (\textit{S4.19}), with the exception of Deepseek, who allowed the use of outputs for training and any other purpose by the user.
All terms mentioned use of automated filtering and detection mechanisms, but did not provide details as to the specific methods used (\textit{S3.16,S4.16}).
The terms of Anthropic, Microsoft, and OpenAI mentioned human involvement in review of inputs and outputs, and those of Anthropic, Google, Microsoft, and OpenAI provided the option of human intervention for suspension or termination decisions (\textit{S6.E}).
\changed{All findings except inputs/outputs being personal data and the use of automated filtering are significant for discussing implications (indicated using red in table).}

\subsection{Applicable Laws \& Consumer Rights}
All service providers except DeepSeek clarified the applicable jurisdiction (\textit{S5.1}) as being the EU and acknowledged the consumer's local laws as applicable (\textit{S5.3}), with Anthropic, Google, and Microsoft specifically mentioning Ireland as the jurisdiction.
DeepSeek was the only service that stated only the laws of China applied regardless of the consumer's location.
Only Mistral mentioned specific laws in the terms (EU's GDPR and AI Act) while other terms referred to laws broadly e.g., consumer laws or vaguely applicable laws (\textit{S5.2}).
Mistral (in France) and DeepSeek (in China) were the only terms that limited where arbitration and legal proceedings could take place (\textit{S5.4}), which affects where users should go to complaint against the provider and where the proceedings will take place.
Though not mentioned in the terms, EU law has specific provisions and legal precedence regarding the jurisdiction of courts within and across Member States, which allows Mistral to utilise the courts in France for all EU consumers \cite{loosWantedBiggerStick2016,CommissionNoticeGuidance2019,SocietaItalianaLastre2025}.

\begin{table}[!ht] 
    \centering
    \caption{Overview of information provided in GenAI service terms regarding applicable jurisdictions and laws. A check mark (\ding{51}) indicates the terms provided this information, and cross mark (\ding{55}) indicates they did not. Colours are used to draw attention to severity of implications.}
    \label{table:findings-legal}
    \small
    \begin{tabularx}{\textwidth}{|p{3cm}|Y|Y|Y|Y|Y|Y|Y|}
    \hline
    \textbf{Topic} & \textbf{Claude (Anthropic)} & \textbf{DeepSeek (Deepseek)} & \textbf{Gemini (Google)} & \textbf{Copilot (Microsoft)} & \textbf{Le Chat (Mistral)} & \textbf{ChatGPT (OpenAI)} \\ \hline
Applied Jurisdictions & \cellcolor{orange!25}\worldflag[width=3mm]{IE}/\worldflag[width=3mm]{EU} & \cellcolor{orange!25}\,\,\worldflag[width=3mm]{CN}* & \cellcolor{orange!25}\worldflag[width=3mm]{IE}/\worldflag[width=3mm]{EU} & \cellcolor{orange!25}\worldflag[width=3mm]{IE}/\worldflag[width=3mm]{EU} & \cellcolor{orange!25}\worldflag[width=3mm]{EU} & \cellcolor{orange!25}\worldflag[width=3mm]{EU} \\ \hline
Local laws apply & \cellcolor{green!25}\ding{51} & \cellcolor{red!25}\ding{55} & \cellcolor{green!25}\ding{51} & \cellcolor{green!25}\ding{51} & \cellcolor{green!25}\ding{51} & \cellcolor{green!25}\ding{51} \\ \hline
Restriction on resolution & \cellcolor{green!25}\ding{55} & \cellcolor{red!25}\,\,\worldflag[width=3mm]{CN}* & \cellcolor{green!25}\ding{55} & \cellcolor{green!25}\ding{55} & \cellcolor{red!25}\,\,\worldflag[width=3mm]{FR}* & \cellcolor{green!25}\ding{55} \\ \hline
Ambiguous restrictions & \cellcolor{red!25}\ding{55} & \cellcolor{red!25}\,\,\ding{55}* & \cellcolor{red!25}\ding{55} & \cellcolor{red!25}\ding{55} & \cellcolor{red!25}\ding{55} & \cellcolor{red!25}\ding{55} \\ \hline
    \end{tabularx}
\end{table}

All providers except Google included statements that referred to vague applicable laws without specifying the exact laws referred to or confirming whether they apply to the current consumer (\textit{S6.D}). 
Anthropic states the user `may have legal rights', while DeepSeek and Mistral include a blanket statement about the terms not affecting any consumer rights.
All terms except Google included a warranty disclaimer that explicitly stated the service is being provided on an `as is' basis without assurances regarding quality, fitness for purpose, accuracy, and reliability. 
Of these, only DeepSeek's disclaimer did not include a statement clarifying its limitation based on applicable law.
The prohibitions provided by Anthropic and OpenAI were phrased as applicable unless restricted by law with others phrased it as applicable to the extent permitted by law. None clarified exactly what was applicable and to what extent.
Of specific interest, only Mistral had requirements regarding the AI Act, according to which the `customer' was prohibited from reporting any `serious incident' to an authority unless required by applicable AI laws, which is supposed to be an obligation for providers and deployers under the AI Act, but the terms did not distinguish this from the individual  consumers.
\changed{The findings regarding applicability and ambiguity of laws are significant for discussing implications (indicated using red in table).}

\subsection{Structure and Accessibility of Terms}
Users were presented with the same information regardless of whether they were free or paid customers or they were individual or enterprise customers (\textit{S1.8}) for all services.
All providers except Mistral had multiple documents that were part of the terms, and additionally  Google and Microsoft had terms that also involved other services and products from their portfolio which made it impossible to determine which terms only applied for use of their GenAI services (\textit{S1.7}).
Additionally, each term contained sections meant for enterprise customers, which needed to be identified and ignored for analysis of consumer sections.
We also faced difficulties in identifying the correct terms for Copilot as the linked URL was redirected to a generic marketing page rather than directly to the terms, and we confirmed this was the case on multiple devices, browsers, and operating systems.
A similar difficulty occurred in the terms for Gemini, which mentioned the possibility for specific terms and pointed to a list which contained `Gemini Apps' which again redirected us back to the same terms.
Only Google explicitly provided a PDF version of the terms, though this did not contain all relevant documents, and Mistral's terms could be exported directed by virtue of being in a single page. For all other terms, we had to take several manual steps to identify and archive each relevant document.

\section{Discussion}\label{sec:discuss}
The terms are a binding legal contract between the GenAI company and the individual consumer.
Existing literature shows terms are not read by users \cite{obarBiggestLieInternet2020}, utilise legal jargon that is difficult to comprehend \cite{robinson2020beyond}, and require regulation to prevent unfair practices and ensure consumer protection \cite{EU_UCTD,EU_UCPD}.
Given that GenAI services are a fairly new phenomenon, we focused our discussions on those aspects which most strongly arise from the nature and use of GenAI as a service. 
We split our discussion into two broad sections - first regarding the complexity and lack of accessibility in terms affecting consumers, and then later discuss implications regarding consumer protection and rights. 
Within each section we also make specific recommendations for GenAI service providers, consumer protection authorities, and policy makers.

\subsection{Implications regarding individual consumers}\label{sec:discuss-user}
In order to sufficiently understand terms and how their use governs GenAI services, we identified the following processes that  
consumers need to undertake: (1) identify all relevant terms; (2) 
comprehend the scope and applicability of the service; (3) understand the roles and responsibilities; (4) formulate expectations regarding the service; and (5) identify controls and know how to operate them. We discuss how the current practices make all of these unrealistic and burdensome even for the most astute consumer, and provide recommendations.

\subsubsection{Lack of necessary information:}
The way terms are structured makes it difficult to identify relevant terms, and differentiating clauses that apply to individual consumers vs  
corporate uses is somewhat a minefield.
This is made even more difficult when terms are complex amalgamations across multiple services, such as those from Google and Microsoft, creating uncertainty and confusion as to which terms apply and what exactly would happen if customers use only the GenAI services.
Additionally, without information and assurances about the specific functionalities, quality expectations, or stability and reliability of the service, a consumer is simply at the mercy of the service provider to accept whatever and however features may be provided without a clear expectation of when service degrades.

In addition to a lack of information, terms also mention that the service may change without prior notice, which could be cosmetic changes such as the user interface, or could be significant changes such as in the underlying models that constitute the bulk of a `GenAI service'.
Even if terms mention a notice period for changes to the terms, the lack of information regarding features and quality metrics means that effectively the consumer must again be at the mercy of the provider to keep the service stable and at the `same' quality.
Despite GenAI developers publishing technical documentation such as model and system cards that can aid in this process, we do not think this is a rational expectation from general consumers.
This means consumers are necessarily burdened to take additional efforts to determine what quality means and to assess it in cases they suspect the service has degraded. 

Considering the nature of GenAI services being marketed as general-purpose tools while also being advertised as being useful for specific cases --- this likely creates an expectation of usefulness for the consumers, who assume the service has been designed to enable such uses.
\changed{Consumer protection law considers marketing and advertised uses as formative towards the decision to start or stop using a service. But what these concepts mean regarding GenAI services is yet to be legally established, especially since our findings show that terms unilaterally claimed no quality assurances and the users sole responsibilities in managing services.
The enforcement of the AI Act's `\textit{intended purpose}' concept \cite{EU_AIAct} also depends on a clear understanding of the advertised uses to delineate between provider and deployer responsibilities.
We therefore point to the need for future studies to specifically investigate the joint applicability of AI Act and consumer protection.}

\begin{recommendations}
(1) GenAI service terms should only pertain to individual consumers and not be mixed with terms for enterprise customers or the provider's other services; 
\\ (2) GenAI providers should provide unambiguous information about the specific features, quality, and stability of the service to inform what the consumer's payments cover
\end{recommendations}

\subsubsection{Unworkable responsibilities put on consumers:}
GenAI terms allocated complete responsibilities and liabilities on the user regarding the inputs, 
in particular assessing 
violation of legal concerns such as copyright and IP. A generous reading of this could consider it 
fair as such inputs are solely decided by users.
However, GenAI services also use automated mechanisms such as filtering and detection of content to determine whether the inputs are valid. These mechanisms  
are `black-boxes' from the user's perspective without clear description  
or ways to control or alter 
them.
In such cases, allocating  
liability  
solely upon users puts them in an impossible position, as they are not in a position to discern which content might trigger violations and risk suspension or termination.

Similarly, for outputs, the responsibilities and liabilities are allocated solely to the user. 
Outputs produced by a GenAI system are a product of the underlying AI model which is determined solely by the company, as well as the input provided by the user.
For any given output that violates a validity requirement, it cannot be stated as a definitive statement that it was produced solely based on the user's input or due to the underlying model.
Further, user inputs are not provided `as is', but are instead combined with more inputs called `system prompts' that instruct the model how respond to the inputs.
Moreover, consumers' only option to stop the violations is to change inputs and hope for the best -- they are not provided any resources, facilities, or guidelines to understand or control how the model produces outputs. 

Providers assigning all liability to the users also means shirking their own role and responsibility, which arises from the degree of control over the model and the service.
This means if incidents such as copyright violations or malicious outputs  
occur, according to the terms, only the user would be solely responsible, regardless of what training data was used by the provider, how the model was developed and configured, and how the service was setup to use the user's input with the model.
A recent example of this is the use of Grok to produce child sexual abuse material (CSAM), where X reiterated that only users are responsible for outputs \cite{belangerBlamesUsersGrokgenerated2026}. 

Invalid inputs and outputs can cause suspension of service and termination of contracts, which are `legal effects' as they impact the legally binding agreement between the consumer and the provider.
Despite this, only 4 providers hinted at human involvement in moderation and final decisions.
Not clarifying whether these decisions are taken entirely algorithmically or there will be a process for complaint and resolution creates uncertainty for users. 
And combined with the lack of information on how the service or mechanisms operate, it also creates a barrier on the consumer to dispute decisions 
as the company is shielded by the terms stating the service was provided without any warranties.

\begin{recommendations}
(1) GenAI providers should provide contractual clarity to consumers on their degree of control over outputs; \\ (2) Consumer protection authorities should investigate the nature of GenAI services and the extent to which consumers can fulfil assigned responsibilities in terms; \\ (3) Changes to regulations may be needed to limit digital services distributing liabilities despite inability of consumers to control service functionalities.
\end{recommendations}

\subsubsection{Asymmetric   
benefits: 
providers  
vs consumers:}
All analysed GenAI services assigned the rights over outputs to the users, gave themselves rights over the submitted inputs, and used inputs and outputs for training regardless of consumers paying for the service, and with the possibility to opt out -- this confirms findings by \citet{leiEffectGenerativeArtificial2025}, who analysed the specifics of the licensing and rights in depth.
While one can argue that training is a trade-off for providing free services, this does not hold when the consumers are made to pay for the service and then to take additional steps to prevent the training.
Prior studies have noted services switching from opt-ins to opt-outs, showing this as a decisive change \cite{king2025user}.
From a consumer protection perspective, this raises concerns about awareness (e.g. whether users are aware of the training and opt-outs), burden (e.g. effort needed to exercise opt-outs), utility (e.g. whether payments create expectation of opt-out), and pricing (e.g. whether opt-outs would cost extra in the future).

All providers except DeepSeek expressly prohibited using outputs for training by the consumers. 
While this restriction can be understood as necessary to avoid other model developers from using the outputs to train their own models, it is not clarified as such and instead is provided in terms for all consumers.
We consider this an unfair trade-off for two reasons: first, the terms explicitly mention that the liability for inputs and outputs is solely assumed by users despite the unworkable imbalance. 
Second, the terms make no distinction between commercial and non-commercial or hobby uses, as well as between free and paid consumers.
If the user is made solely liable and responsible, but then restricted from using the outputs as they deem fit and in a manner that does not economically harm the provider, then such practices present an 
imbalance and unfair  
restrictions on 
the consumer.

This same pattern is repeated 
regarding reverse engineering, where specific prohibitions are mentioned with the caveat that applicable laws may change the nature of such prohibitions by limiting them or removing them entirely.
We do not expect  
a typical consumer to be aware of what exactly the laws state, 
nor expect the consumers to  
be aware of what specific laws would be applicable here.
Such statements, therefore,  
create an environment where users are expected to figure out what the applicable laws are and whether a given restriction or prohibition is applicable to them.
Unlike companies, it is unlikely for consumers to have a lawyer on retainer who can help figure out which laws the terms are referring to.
This can also lead to cases where users may believe that the restrictions apply, restricting their ability to fully enjoy the service or exercise their legal rights.

\begin{recommendations}
(1) GenAI Providers use of data (e.g. for training) should mean they have liability for it;  
\\ (2) GenAI Providers should not restrict users from using outputs under their responsibility; \\ (3) Terms should provide absolute clarity on whether a clause applies or not, and should not burden consumers with legal investigations beyond their ability.
\end{recommendations}

\subsection{Implications under EU Consumer Protection Law and Consumer Rights}\label{sec:discuss-legal}
Consumer protection law in EU is a complex multi-layered framework comprising of key directives\footnote{For our discussion, we consider the Directives as providing a harmonised law across EU and do not consider implementing laws in Member States.} regarding consumer rights \cite{EU_consumer-rights-directive} and digital services contracts \cite{EU_contract_digital_service_directive} -- which shape the content provided in terms, with Unfair Contract Terms Directive (UCTD) \cite{EU_UCTD} and Unfair Commercial Practices Directive (UCPD) \cite{EU_UCPD} ensuring fairness.
UCTD Article 3(1) defines `unfair terms' based on three criteria: (1) contrary to the requirement of `\textit{good faith}', (2) causes a `\textit{significant imbalance}' in  rights and obligations, and (3) acts to the \textit{detriment of the consumer}.
UCPD Article 5(4) defines `\textit{unfair practices}' as based on being misleading or aggressive.
Both UCTD and UCPD provide an annex that lists specific terms and practices that \changed{`may be unfair' (i.e. potentially unfair)}, with the UCPD prohibiting specific practices as always being unfair.
We discuss the implications of our findings based on the potential applicability of these two provisions.
In doing so, we bring to attention the extent to which  
the terms and the accompanying practices associated with GenAI services require further exploration under consumer protection laws.
To inform our arguments, we consulted authoritative guidance by a consumer protection authority \cite{GuideUnfairTerms} and the rationale for unfair practices being regulated \cite{CommissionNoticeGuidance2019}. 

\subsubsection{Good faith in terms}

`Good faith' requires not taking advantage of the consumer's lack of experience and weak bargaining position \cite{GuideUnfairTerms,CommissionNoticeGuidance2019}.
In our findings, all providers required specific restrictions and prohibitions to be fulfilled with complete responsibility and liabilities assumed solely by consumers.
However, as we pointed out in the previous sections, 
these terms are difficult to decipher for a typical consumer. 
We also argued that 
the lack of quality metrics in terms, for example to note that accuracy or reliability could be measured, means that consumers have no way of understanding whether the outputs they receive as part of the service are problematic or acceptable.
While information related to these matters could have been provided in other documentation outside the terms (which we did not assess), their lack of acknowledgement within the terms  
creates a burden for consumers who would have to discover them through other means.
And in case of disputes, require additional efforts as the information was not part of the terms and could have been changed without their knowledge.

Good faith  
requires open dealing, which requires that terms be provided in plain intelligible language that an average consumer can understand, available in a single location, and expressed legibly with no hidden small print. 
Our findings showed these documents are burdensome to access and understand  
where 5 out of 6 providers  
 used multiple pages that could only be found by reading the terms carefully due to being mixed with terms regarding enterprise customers and provider's other services.
We also found significant hurdles in discerning 
language used to outline  
restrictions, warranties, disclaimers, and similar legal provisions despite these being one of the most important clauses for consumers \cite{liuIdentifyingTermsConditions}. The complex phrasing of these clauses without clarification of impacts also likely violates UCTD requiring `plain intelligible language' that consumers understand.

\begin{recommendations}
(1) GenAI Providers should ensure terms are compatible with the spirit and letter of EU consumer law by making them intelligible for consumers, providing all relevant terms together, and avoid contractual obligations on consumers that they are not equipped to fulfil; \\ (2) Consumer protection authorities should assess whether the terms follow their respective good faith interpretations and issue targeted guidance for GenAI.
\end{recommendations}

\subsubsection{Imbalance in terms}
For imbalance arising from terms to be considered unfair, there must be a significant advantage to the provider without providing an equal benefit to a consumer.
As can be observed from our findings, a clear imbalance exists in the allocation of liabilities, where the consumer is made solely liable for the use of the service without assurances 
to its quality, fitness, or purpose. 
Imbalance also exists in terms that enable training of models by the provider without providing this benefit to consumers, increased by requiring consumers to opt-out, and further in cases where data may still be used even after opting out.
Imbalance may also be present in cases where the service provider, by virtue of being the model developer, has specific access to information and capabilities to direct the model towards specific behaviours.

This creates a situation where the consumer must ensure that their inputs and outputs meet the validity as determined solely by the service provider, with the only mechanism available to them as changing inputs, while the service provider is in a stronger position by virtue of having complete control over the underlying model as well as peripheral measures such as filtering and detection.
Combined with the earlier arguments regarding disproportionate allocation of responsibilities and liabilities, it is evident that the consumer takes on more liabilities in the use of the service than the service provider does so in its provision of the same service.

\begin{recommendations}
(1) GenAI Providers should redraft their terms to ensure a balanced implementation of rights and obligations without detriment to consumers -- as required by EU law; \\ (2) Consumer protection authorities and policy makers should consider introducing additional terms and practices as unfair to tackle the specific challenges of responsibilities and balances within GenAI services, including potential rights regarding training opt-outs and restrictions on use of outputs.
\end{recommendations}

\subsubsection{Potentially unfair terms}
The disclaimers used in terms indicate services have no assurance of quality, performance, accuracy, stability, or fitness of purpose, which brings into question what the consumers are paying for and if they have no options but to accept any and all deficiencies and errors when using the service.
Yet, the hype for GenAI and their use in services is increasing exponentially~\cite{Bender-Hanna-2025,varoquaux2025hypesustainabilitypricebiggerisbetter}.
Only one of two things therefore must be true: either these services or their providers are not capable of achieving sufficiently good qualities, or that the providers use these statements to avoid responsibilities that would be due to them under EU consumer protection laws \cite{EU_consumer-rights-directive,EU_contract_digital_service_directive}.

In the previous two sections, we identified the potential for specific terms and practices to be considered unfair based on the principles and definitions  
outlined in UCTD \cite{EU_UCTD} and UCPD \cite{EU_UCPD} laws.
In this section, we specifically explore the relation between these findings and the lists provided in both laws regarding unfair and prohibited practices.
UCTD Annex I(1)(b) states ``\textit{excluding or limiting the legal rights of the consumer [...] in the event of total or partial non-performance or inadequate performance}'' as a `potentially unfair' term. 
This matches our findings where no terms included specific quality metrics that would enable consumers to detect and dispute performance.
In particular, Deepseek's assertion that only the laws of China would apply seems to be a clear exclusion of the legal rights of a consumer based in EU, for example to ensure fairness in contracts and regarding disputes.
It also prevents consumers in EU from taking action against Deepseek without opening proceedings in China, which is contrary to EU consumer protection norms and incompatible with established legal precedents \cite{loosWantedBiggerStick2016,CommissionNoticeGuidance2019,SocietaItalianaLastre2025}. 
This clause and its use by Deepseek is thus \textit{unfair} within the meaning of UCTD, and in addition also impacts the fundamental right of consumer protection \cite{EU_CFR}.

Another instance is UCTD Annex I(1)(k) which states ``enabling the seller or supplier to alter unilaterally without a valid reason any characteristics of the product or service to be provided'' as a potentially unfair term.
This matches our findings where no terms provided features or functionalities associated with the service, did not provide assurances regarding quality, and indicated that the service may change at the sole discretion of the provider and without prior notice to the consumers.
The categorisation of these findings as being unfair depends on the justification provided by companies to alter the services, which typically tends to be associated with improvement to services or security.
For this we should consider that the service is intended for and used by the consumer, and the validity of such benefits should therefore rest on whether the benefits from changes to the service are intended or materialise for the consumers, and that the consumers are informed thereof.
We should also consider whether the consumer has the ability to direct the change or control when it happens, which our findings show is not provided for any of the GenAI services.
These practices therefore also are potentially \textit{unfair}.

In continuation, UCTD Annex I(1)(m) notes that provider having the sole right to determine whether services are in conformity with the terms may be considered unfair.
Since the terms do not give the consumer any information or capability to determine the quality of the service, and instead disclaim all assurances explicitly, the consumer may be at a significant disadvantage as the terms may be used to consider any defects or issues regarding reliability, performance, and use may be considered as non-applicable by the service provider. 
Additionally, even if the terms state they would change with a notice period, the service itself changing without prior notice or control can be seen as a change of terms \cite{loosWantedBiggerStick2016}. 
Similarly, providers determining quality unilaterally without awareness of consumers may also be seen as a change of terms \cite{loosWantedBiggerStick2016}.
Combined with the imbalances discussed in prior sections, this is potentially unfair as the consumer must rely solely on the service provider to determine whether the service provided is of a sufficient quality.
Further explorations would require an analysis of the marketing and information provided to consumers prior to purchase/use as well as the documentation provided in context of the service, with analysis based on UCPD Articles 6 and 7 regarding misleading actions and omissions respectively.

\begin{recommendations}
(1) Our findings have shown that current GenAI terms contain potentially unfair clauses and practices as defined in UCTD and UCPD. This provides sufficient ground for Consumer protection authorities to investigate further; \\ (2) Academic and other researchers should conduct scientific analysis of GenAI provider's marketing and practices to assess implied capabilities against their total disclaim in terms.
\end{recommendations}

\subsection{Limitations of this work}
\changed{
(1) The scope of our work only considered the terms associated with GenAI services. It is highly likely that some issues we have highlighted, such as the use of warranty disclaimers, is a widespread practice. We believe our codebook would be useful in identifying and analysing this via a longitudinal study.
(2) We only considered the information provided within the terms document as determined by our selection criteria. The existence of other forms of documentation not linked to the terms may contain information of relevance, though we also consider this as a burden on the consumer to expend effort to identify. Our focus on terms fills an important gap as past work has only looked at privacy policies without analysing the contractual and consumer protection aspects of GenAI services. 
(3) We also did not consider the information provided to consumers via advertisements, marketing such as on websites, and the specific information given via the service interfaces. A study of these would be of relevance to better understand the extent of information provided to consumers and further discussions on the severity of our findings under consumer protection.
(4) We only discussed our findings in context of context protection, though we have highlighted the relevance and necessity of further analysis under the GDPR and the AI Act, and potentially their combined enforcement. We also did not explore the relevance to the Digital Services Act (DSA) and Digital Markets Act (DMA), especially with Google (Alphabet) and Microsoft as highly regulated platforms and `gatekeepers'.
(5) We did not consider variance exist EU within Member States -- such as regarding age requirements and consumer protection. Additionally, our findings also have relevance to other jurisdictions based on similarity of terms and laws (or their absences). These would require modifications to the codebook and the analysis.
}

\section{Conclusions}\label{sec:conclusion}
Our findings show how the legally binding terms of GenAI services state that they are not fit for any purpose, cannot be provided with any assurance of quality, will not be accurate or error-free, and a myriad of qualities that would be reasonably concluded as `problems' rather than functionalities. 
Our findings also show how the terms pass all responsibilities on to the user, even for cases where the user has no ability to fulfil them -- while providers allow themselves to benefit from the data and labour provided by users for training their models.
Finally, we highlight the myriad issues and imbalances faced by consumers, who instead of benefiting from robust legal protections, actually suffer from the complexities and autocracy imposed by the terms of popular AI tools. 
We also raise urgent concerns regarding GenAI services and their dilution of consumer rights, freedoms, and ability to enjoy the services without undue liabilities.
Rather than being isolated instances, these practices have become systemic, as evidenced from past work that has highlighted similar issues in big providers \cite{loosWantedBiggerStick2016} and the growing trend of companies basing their terms on one another \cite{edwardsPrivateOrderingGenerative2025}.
The specific issues we identified and our perspectives in analysing them and developing recommendations are rooted in the areas of consumer protection, which is a fundamental right within the EU~ \cite{EU_CFR}.
We are convinced that advances in technologies, including GenAI, must thus be done in a manner that respects this right, rather then discarding  it, especially given the increasingly prevalent usage of AI technologies in different tools and applications. 
\changed{For this reason, we have submitted our preliminary findings \cite{consultation_2025_DFA} to the proposed Digital Fairness Act which is expected to significantly update consumer protection regulation(s) in EU, and plan to continue our engagement in its next stages later this year. We also plan to share our work and work with civil society organisations and  authorities to develop additional guidance on regulating GenAI services -- especially to harmonise enforcement under multiple regulatory frameworks.}

\changed{\subsubsection*{Updates:} Several GenAI services (Claude, Le Chat, ChatGPT) updated their terms, usage policies, or privacy policies after 20th November 2025 when our analysis and findings were completed. While we do not analyse their updated terms in this paper, our preliminary analysis shows that several of the important issues we have highlighted in this work continue to persist. Please see \underline{\url{https://aial.ie/research/terms-of-abuse}} for further updates.}

\begin{acks}
The AI Accountability Lab is supported by grants from the John D. and Catherine T. MacArthur Foundation, the AI Collaborative of the Omidyar Group, Luminate Foundation, European AI \& Society Fund, and Bestseller Foundation.
\end{acks}

\section*{Generative AI Usage Statement}

The author(s) did not use generative AI tools during manuscript preparation.

\bibliographystyle{ACM-Reference-Format}
\bibliography{ref}

\appendix
\section{Analysed Terms}\label{appendix:terms}
\begin{enumerate}
    \item Claude (Anthropic) \url{https://claude.ai/}
    \begin{enumerate}[label=\theenumi.\arabic*]
        \item Consumer Terms \url{https://www.anthropic.com/legal/consumer-terms} (dated 8 October 2025)
        \item Privacy Policy \url{https://www.anthropic.com/legal/privacy}
        \item Usage Policy \url{https://www.anthropic.com/legal/aup}
    \end{enumerate}
    \item DeepSeek \url{https://chat.deepseek.com}
    \begin{enumerate}[label=\theenumi.\arabic*]
        \item Terms of Use \url{https://cdn.deepseek.com/policies/en-US/deepseek-terms-of-use.html} (dated 28 April 2025)
        \item Privacy Policy \url{https://cdn.deepseek.com/policies/en-US/deepseek-privacy-policy.html}
        \item Open Platform Terms of Service \url{https://cdn.deepseek.com/policies/en-US/deepseek-open-platform-terms-of-service.html}
    \end{enumerate}
    \item Gemini (Google) \url{https://gemini.google.com}
    \begin{enumerate}[label=\theenumi.\arabic*]
        \item Policies and Terms \url{https://policies.google.com/} which provides link to terms
        \item Terms of Service \url{https://policies.google.com/terms} (dated 22 May 2024)
        \item Generative AI-Prohibited Use Policy \url{https://policies.google.com/terms/generative-ai/use-policy}
        \item Service-specific additional terms and policies \url{https://policies.google.com/terms/service-specific} which contains entry for `Gemini Apps' and provides same links to Terms and Generative AI Policy as above
    \end{enumerate}
    \item Copilot (Microsoft) \url{https://www.copilot.com/}
    \begin{enumerate}[label=\theenumi.\arabic*]
        \item `terms' link \url{https://www.bing.com/new/termsofuse?utm_source=copilot.com} is redirected to landing page \url{https://www.microsoft.com/en-ie/microsoft-copilot/for-individuals?form=MA13YT}. On this page, the link to `Copilot Terms' \url{https://aka.ms/consumercopilotterms} redirects to the same landing page. Another link `Terms of Use' \url{https://go.microsoft.com/fwlink/?LinkID=206977} redirects to terms \url{https://www.microsoft.com/en-us/legal/terms-of-use} (dated 2 February 2022)
        \item Privacy Statement \url{http://go.microsoft.com/fwlink/?linkid=248681} which is redirected to \url{https://www.microsoft.com/en-gb/privacy/privacystatement}
        \item Services Agreement \url{https://www.microsoft.com/en-us/servicesagreement}
        \item Bing Image Creator and Bing Video Creator Terms of Use \url{https://www.bing.com/new/termsofuseimagecreator}
        \item Web search (using Bing) led to `Copilot Terms of Use' \url{https://www.microsoft.com/en-us/microsoft-copilot/for-individuals/termsofuse} which were not linked in any of the above pages
    \end{enumerate}
    \item Le Chat (Mistral) \url{https://chat.mistral.ai/}
    \begin{enumerate}[label=\theenumi.\arabic*]
        \item Legal terms and conditions \url{https://mistral.ai/terms/#terms-of-service} which redirects to \url{https://legal.mistral.ai/terms} (dated 27 May 2025)
    \end{enumerate}
    \item ChatGPT (OpenAI) \url{https://chatgpt.com/}
    \begin{enumerate}[label=\theenumi.\arabic*]
        \item Terms \& Policies \url{https://openai.com/policies/} which provides links to Europe Terms of Use \url{https://openai.com/policies/terms-of-use/} (dated 29 April 2025)
        \item Service Terms \url{https://openai.com/policies/service-terms/}
        \item Usage Policies \url{https://openai.com/policies/usage-policies/}
        \item Sharing \& Publication Policy \url{https://openai.com/policies/sharing-publication-policy/}
        \item Service Credit Terms \url{https://openai.com/policies/service-credit-terms/}
        \item Transparency \& Content Moderation \url{https://openai.com/transparency-and-content-moderation/}
        \item How your data is used to improve model performance \url{https://openai.com/policies/how-your-data-is-used-to-improve-model-performance/}
        \item Privacy Policy \url{https://openai.com/policies/eu-privacy-policy/} was linked but not included for analysis as the terms stated: ``Although it does not form part of these Terms...''
    \end{enumerate}
\end{enumerate}

\section{Codebook}\label{appendix:codebook}

The codebook fields each have an assigned ID, Field/Question, columns `Info on main page' and `Info on secondary page' to document whether information was found in the terms document other (secondary) linked documents, Notes for annotator to record observations, and Example provide indicative phrases that may be used in terms to find this information. The two info columns contain an enumerated list of options for most questions from which annotators select the most applicable option if present, and use the notes column to record additional information.

\begin{enumerate}[label=S\theenumi]
    \item Metadata
    \begin{enumerate}[label=S1.\arabic*]
        \item Timestamp of annotation
        \item Annotator identifier
        \item Target
        \item Link to archived terms (folder)
        \item Additional Links found during annotation
        \item Terms structure
        \begin{enumerate}[label=\alph*]
            \item Single page / monolithic
            \item Paginated / multiple pages
            \item Single page with links to other non-terms pages
        \end{enumerate}
        \item Same terms used for free or paid services? \textit{(Sample: If you are a free user…; If you have an account with us)}
        \begin{enumerate}[label=\alph*]
            \item Terms apply for free services
            \item Terms apply for paid services
            \item Same terms apply for free and paid services
        \end{enumerate}
    \end{enumerate}
    \item Service Provision
    \begin{enumerate}[label=S2.\arabic*]
        \item Is what constitutes as the ‘Service’ described with a specific name? \textit{Sample: These terms relate to ChatGPT, Service is defined as xyz}
        \begin{enumerate}[label=\alph*]
            \item Yes, the Service is described + text field
            \item No, the Service is not described
        \end{enumerate}
        \item If the Service is described, are specific features or functionalities described as part of the Service?             \textit{Sample: By service we mean the ability to use the model; Features such as image generation}
        \begin{enumerate}[label=\alph*]
            \item Yes, the Service is described + text field
            \item No, the Service is not described
        \end{enumerate}
        \item Do the terms mention how user will be informed and can control changes to the service?              \textit{Sample: We may change the Service at our discretion without prior notice or warning; Where we change our model, we will allow users to opt-in to the model}
        \begin{enumerate}[label=\alph*]
            \item Yes, the terms explicitly mention that the Service will not change unless the User enables it
            \item No, the terms explicitly mention that the Service will change and the User will be given a notice period
            \item No, the terms explicitly mention that the Service will change but do not state how the User will be informed about this
            \item No, the terms do not state whether the Service will change
        \end{enumerate}
        \item Do the terms describe the Service in terms of specific quality?             \textit{Sample: We have a 99\% uptime; Our models are guaranteed to provide correct output 99\% of the time   Note: Mentioning the model hallucinates is NOT a quality metric unless it is accompanied with information on when/how much}
        \begin{enumerate}[label=\alph*]
            \item Yes, quality metrics are described regarding (e.g. speed of responses)
            \item Yes, quality metrics are described regarding accuracy (e.g. accuracy of responses)
            \item Yes, quality metrics are described regarding availability (e.g. uptime)
            \item Yes, other quality metrics are provided + text
            \item No, quality metrics are not provided
        \end{enumerate}
        \item If quality metrics are described, do the terms provide any assurances or claims or guarantees regarding them?               \textit{Sample: We guarantee xyz with \%, We will provide xyz with \%}
        \begin{enumerate}[label=\alph*]
            \item Yes, assurances or guarantee is mentioned
            \item No, assurances or guarantee is not mentioned
        \end{enumerate}
    \end{enumerate}
    \item Service Usage - Inputs
    \begin{enumerate}[label=S3.\arabic*]
        \item[S6.A] Combined concept for both inputs and outputs?
        \item[S6.B] Are inputs and outputs declared as personal data?
        \item What forms or modalities can the User provide their input as?               \textit{Sample: You can provide input through the service or by uploading a picture}
        \begin{enumerate}[label=\alph*]
            \item Text
            \item Images
            \item Videos
            \item Audio
            \item Not mentioned 
        \end{enumerate}
        \item What sources can the User provide their input as?
        \begin{enumerate}[label=\alph*]
            \item Directly entering it into the Service (e.g. typing, uploading)
            \item Through the use of their Device (e.g. Camera, GPS)
            \item Third Party (e.g. Siri)
            \item Not mentioned
        \end{enumerate}
        \item Are there restrictions regarding what input is exclusively permitted?               \textit{Sample: We only allow you to upload; We only accept input}
        \begin{enumerate}[label=\alph*]
            \item Yes, the terms mention  exclusively what input is permitted + text
            \item “Everything is permitted” is explicitly mentioned
            \item Not mentioned
        \end{enumerate}
        \item Are there restrictions regarding what input is prohibited?              \textit{Sample: We prohibit content where; We do not accept input}
        \begin{enumerate}[label=\alph*]
            \item Yes, the terms mention prohibited input + text
            \item “No prohibitions” is explicitly mentioned
            \item Not mentioned
        \end{enumerate}
        \item[S6.C]	Is reverse engineering explicitly prohibitied?
        \item Are there restrictions regarding the scope of input such that the Service is only described as intended for the scope?
        \begin{enumerate}[label=\alph*]
            \item Yes, specific input categories are described as the scope + text
            \item Nothing is out of scope is explicitly mentioned
            \item Not mentioned
        \end{enumerate}
        \item How do the terms describe the ownership and retention of rights regarding the provided User input?
        \begin{enumerate}[label=\alph*]
            \item User retains ownership, and Service provider is granted rights only for the Service
            \item User retains ownership, and Service provider is granted rights for the Service as well as any other use
            \item Rights are transferred from the User to the Service Provider
            \item Not mentioned
        \end{enumerate}
        \item How do the terms describe the use of User input for further training or refinement of Model by the service provider?
        \begin{enumerate}[label=\alph*]
            \item Inputs will be used for training/refinement
            \item Inputs will not be used for training/refinement
            \item Not mentioned
        \end{enumerate}
        \item If User input will be used for further training or refinement of the Model, what options or controls does the User 
        have?
        \begin{enumerate}[label=\alph*]
            \item User has no controls
            \item User must opt-in and information is provided on how to do this
            \item User must opt-in but no information is provided on how to do this
            \item User must opt-out and information is provided on how to do this
            \item User must opt-out but no information is provided on how to do this
            \item Not mentioned
        \end{enumerate}
        \item Will the User input be used for other purposes beyond training or refinement of the Model?
        \begin{enumerate}[label=\alph*]
            \item Inputs will be used for analysis or measurements regarding the Service
            \item Inputs will be used for research and product development
            \item Other + text
            \item Not mentioned
        \end{enumerate}
        \item Will the User input be shared with Third Parties?
        \begin{enumerate}[label=\alph*]
            \item Yes
            \item No
            \item Not mentioned
        \end{enumerate}
        \item If User input will be shared with Third Parties, are these third parties identified?
        \begin{enumerate}[label=\alph*]
            \item Yes, the identities of Third Parties are provided
            \item Yes, the categories of Third Parties are provided
            \item Not mentioned
        \end{enumerate}
        \item If User input will be shared with Third Parties, are the specific purposes for why they will be shared mentioned?
        \begin{enumerate}[label=\alph*]
            \item Yes + text
            \item Not mentioned
        \end{enumerate}
        \item If User input will be shared with Third Parties, will it be in a privacy preserving form?
        \begin{enumerate}[label=\alph*]
            \item Yes, and specific measures are provided + text
            \item Yes, but specific measures are not provided
            \item Not mentioned
        \end{enumerate}
        \item When submitting input, what responsibilities are allocated to the User regarding the validity of inputs?
        \begin{enumerate}[label=\alph*]
            \item Copyright violation does not occur
            \item Input meets safety standards and these are mentioned + text
            \item Input meets safety standards but these not mentioned
            \item Other + text
            \item Not mentioned
        \end{enumerate}
        \item If the input does not meet the validity requirements, is the resulting liability explicitly clarified?
        \begin{enumerate}[label=\alph*]
            \item Yes, User assumes liability
            \item Yes, User and Service share liability
            \item Yes, Service assumes liability
            \item Not mentioned
        \end{enumerate}
        \item Does the service provider use filtering or detection mechanisms over user input?
        \begin{enumerate}[label=\alph*]
            \item Yes, and specific measures are provided + text
            \item Yes, but specific measures are not provided
            \item Not mentioned
        \end{enumerate}
        \item What happens if the filtering/detection mechanism detects a violation or problem regarding the user input?
        \begin{enumerate}[label=\alph*]
            \item User may lose access to the Service
            \item User assumes liability
            \item Other + text
            \item Not mentioned
        \end{enumerate}
    \end{enumerate}
    \item Service Usage - Outputs
    \begin{enumerate}[label=S4.\arabic*]
        \item What forms or modalities will the output be provided as?
        \begin{enumerate}[label=\alph*]
            \item Text
            \item Images
            \item Videos
            \item Audio
            \item Not mentioned
        \end{enumerate}
        \item How can the User access the output?
        \begin{enumerate}[label=\alph*]
            \item Directly through the Service (e.g. see it on the website)
            \item Third Party (e.g. Siri)
            \item Publish (e.g. public access via Service)
            \item Not mentioned
        \end{enumerate}
        \item Are there restrictions regarding what output is exclusively permitted?              \textit{Sample: We will only provide; We only generate output}
        \begin{enumerate}[label=\alph*]
            \item Yes, the terms mention  exclusively what output is permitted + text
            \item “Everything is permitted” is explicitly mentioned
            \item Not mentioned
        \end{enumerate}
        \item Are there restrictions regarding what output is prohibited?             \textit{Sample: We prohibit content where; We do not produce output}
        \begin{enumerate}[label=\alph*]
            \item Yes, the terms mention prohibited output + text
            \item “No prohibitions” is explicitly mentioned
            \item Not mentioned
        \end{enumerate}
        \item Are there restrictions regarding the scope of output such that the Service is only described as intended for the scope?
        \begin{enumerate}[label=\alph*]
            \item Yes, specific output categories are described as the scope + text
            \item Nothing is out of scope is explicitly mentioned
            \item Not mentioned
        \end{enumerate}
        \item How do the terms describe the ownership and retention of rights regarding the output?
        \begin{enumerate}[label=\alph*]
            \item User retains ownership with no information about rights of Service Provider
            \item User retains ownership, and Service provider is granted rights for the Service as well as any other use
            \item Rights are retained by the Service Provider
            \item Not mentioned
        \end{enumerate}
        \item How do the terms describe the use of output for further training or refinement of Model by the Service Provider?
        \begin{enumerate}[label=\alph*]
            \item Outputs will be used for training/refinement
            \item Outputs will not be used for training/refinement
            \item Not mentioned
        \end{enumerate}
        \item If Output will be used for further training or refinement of the Model, what options or controls does the User have?
        \begin{enumerate}[label=\alph*]
            \item User has no controls
            \item User must opt-in and information is provided on how to do this
            \item User must opt-in but no information is provided on how to do this
            \item User must opt-out and information is provided on how to do this
            \item User must opt-out but no information is provided on how to do this
            \item Not mentioned
        \end{enumerate}
        \item Will the Output be used for other purposes beyond training or refinement of the Model?
        \begin{enumerate}[label=\alph*]
            \item Outputs will be used for analysis or measurements regarding the Service
            \item Outputs will be used for research and product development
            \item Other + text
            \item Not mentioned
        \end{enumerate}
        \item Will the Output be shared with Third Parties?
        \begin{enumerate}[label=\alph*]
            \item Yes
            \item No
            \item Not mentioned
        \end{enumerate}
        \item If Output will be shared with Third Parties, are these third parties identified?
        \begin{enumerate}[label=\alph*]
            \item Yes, the identities of Third Parties are provided
            \item Yes, the categories of Third Parties are provided
            \item Not mentioned
        \end{enumerate}
        \item If Output will be shared with Third Parties, are the specific purposes for why they will be shared mentioned?
        \begin{enumerate}[label=\alph*]
            \item Yes + text
            \item Not mentioned
        \end{enumerate}
        \item If Output will be shared with Third Parties, will it be in a privacy preserving form?
        \begin{enumerate}[label=\alph*]
            \item Yes, and specific measures are provided + text
            \item Yes, but specific measures are not provided
            \item Not mentioned
        \end{enumerate}
        \item When generating output, what responsibilities are allocated to the User regarding the validity of outputs?
        \begin{enumerate}[label=\alph*]
            \item Copyright violation does not occur
            \item Output meets safety standards and these are mentioned + text
            \item Output meets safety standards but these not mentioned
            \item Other + text
            \item Not mentioned
        \end{enumerate}
        \item If the output does not meet the validity requirements, is the resulting liability explicitly clarified?
        \begin{enumerate}[label=\alph*]
            \item Yes, User assumes liability
            \item Yes, User and Service share liability
            \item Yes, Service assumes liability
            \item Not mentioned
        \end{enumerate}
        \item  Will the generated Output be subject to any form of filtering or detection?
        \begin{enumerate}[label=\alph*]
            \item Yes and details are provided + text
            \item Yes but details are not provided
            \item Not mentioned
        \end{enumerate}
        \item What happens if the filtering/detection mechanism detects a violation or problem regarding the outputs?
        \begin{enumerate}[label=\alph*]
            \item User may lose access to the Service
            \item User assumes liability
            \item Other + text
            \item Not mentioned
        \end{enumerate}
        \item[S6.E]	For automated methods related to filtering and detection of validity, will the decision involve human oversight or confirmation, and will the user have the ability to request human review of the decision?
        \item Are specific ‘risks’ or ‘issues’ identified regarding the output that would reduce the quality or cause detriments?
        \begin{enumerate}[label=\alph*]
            \item Inaccuracy
            \item Bias
            \item Hallucination
            \item Other + text
            \item Not mentioned
        \end{enumerate}
        \item Are there specific restrictions on the use of outputs for training or refinement of outside of the Service?
        \begin{enumerate}[label=\alph*]
            \item No restrictions
            \item User can reuse outputs for training or refinement
            \item User cannot reuse outputs for training or refinement
            \item Not mentioned
        \end{enumerate}
        \item Do the terms inform or warn the User regarding specific harms that may arise from the use of the service ?
        \begin{enumerate}[label=\alph*]
            \item Yes, regarding risks to safety (e.g. mental health)
            \item Yes, regarding risks of the Service (e.g. inaccurate, unreliable)
            \item Yes + text
            \item Not mentioned
        \end{enumerate}
    \end{enumerate}
    \item Legal
    \begin{enumerate}[label=S4.\arabic*]
        \item Is the use of the Service restricted to a specific jurisdiction?                \textit{E.g. EU/EEA}
        \begin{enumerate}[label=\alph*]
            \item Yes, to EU
            \item Yes, to USA
            \item Yes, to China
            \item No restrictions
            \item Other + text
            \item Not mentioned
        \end{enumerate}
        \item Are specific laws mentioned governing the Service?            \textit{  E.g. GDPR}
        \begin{enumerate}[label=\alph*]
            \item Yes, GDPR and other EU laws
            \item Other + text
            \item Not mentioned
        \end{enumerate}
        \item Does the User’s location or jurisdiction affect the laws applicable to the Service?
        \begin{enumerate}[label=\alph*]
            \item Yes, laws in User’s jurisdiction apply
            \item No, only those jurisdictions mentioned in the terms apply
            \item Not mentioned
        \end{enumerate}
        \item Are there restrictions on where arbitration or legal proceedings can take place?
        \begin{enumerate}[label=\alph*]
            \item Dispute resolution + text
            \item Arbitration + text
            \item Enforcement via courts + text
            \item Other + text
        \end{enumerate}
        \item[S6.D]	Do the terms explicitly include statements that create exceptions based on applicable law without specifiying which laws exist and whether they are applicable for the user in this context?
    \end{enumerate}
\end{enumerate}

\section{Findings}\label{appendix:findings}
\textit{(see next page)}
\begin{table}[!ht]\label{table:appendix-findings} 
    \centering
    \caption{Overview of findings from investigating terms of GenAI services using the developed codebook. Table only shows if a topic was present, with further details provided in the accompanying text. Asterisks (*) indicate special case or additional information}
    \begin{tabular}{|p{1.25cm}|p{2.5cm}|l|l|l|l|l|l|}
    \hline
        \textbf{ID} & \textbf{Topic} & \textbf{Claude} & \textbf{DeepSeek} & \textbf{Gemini} & \textbf{Copilot} & \textbf{Le Chat} & \textbf{ChatGPT} \\ \hline
        ~ & \textbf{Company} & \textbf{Anthropic} & \textbf{DeepSeek} & \textbf{Google} & \textbf{Microsoft} & \textbf{Mistral} & \textbf{OpenAI} \\ \hline
        S1.7 & documents & 3 & 3 & 2 & 5 & 1 & 7 \\ \hline
        S1.8 & free vs paid & same & same & same & same & same & same \\ \hline
        S2.4-S2.5 & QA & no & no & no & no & no & no \\ \hline
        S6.A & I/O label & Materials & n/a & Content & Content & Data & Content \\ \hline
        S6.B & personal data? & yes & yes & yes & yes & yes & yes \\ \hline
        S3.3-S3.5, S4.3-S4.5 & I/O restrictions & yes & yes & yes & yes & yes & yes \\ \hline
        S6.C & reverse engineering allowed? & no* & no* & no* & no* & no* & no* \\ \hline
        S3.6 & input rights retained & yes & yes & yes & yes & yes & yes \\ \hline
        S4.6 & output rights given & yes & yes & yes & yes & yes & yes \\ \hline
        S3.7, S4.7 & training by provider & yes & yes & yes* & yes & yes & yes \\ \hline
        S3.8, S4.8 & user control & opt-out & opt-out & opt-out* & opt-out & opt-out & opt-out \\ \hline
        S4.19 & training by user & no & yes & no & no & no & no \\ \hline
        S3.9, S4.9 & other purposes & yes & yes & yes & yes & yes & yes \\ \hline
        S3.10-S3.13, S4.10-S4.13 & third party sharing & yes & yes & yes & yes & yes & yes \\ \hline
        S3.14 & input liability & user & user & user & user & user & user \\ \hline
        S4.14 & output liability & user & user & user & user & user & user \\ \hline
        S3.16, S4.16 & filtering/detection & yes & yes & yes & yes & yes & yes \\ \hline
        S3.17, S4.17 & violation causes suspension or termination & yes* & yes & yes & yes & yes & yes \\ \hline
        S6.E & human involvement & yes & n/a & yes* & yes & n/a & yes \\ \hline
        S4.18 & technical risks & yes & yes & yes & yes & yes & yes \\ \hline
        S4.20 & harms to user & n/a & n/a & n/a & n/a & n/a & n/a \\ \hline
        S5.1 & applicable jurisdiction & Ireland/EU & China* & Ireland/EU & Ireland/EU & EU* & EU* \\ \hline
        S5.2 & mentioned laws & n/a & n/a & n/a & n/a & GDPR, AI Act & n/a \\ \hline
        S5.3 & local laws apply & yes & no & yes & yes & yes & yes \\ \hline
        S5.4 & restriction on courts & no & China* & no & no & France* & no \\ \hline
        S6.D & restrictions unless laws apply & yes & n/a* & n/a & yes & yes & yes \\ \hline
    \end{tabular}
\end{table}

\end{document}